\documentstyle[12pt]{article}
\newcommand{\reff}[1]{(\ref{#1})}
\def\reff #1{(\ref{#1})}
\def\bra #1{\langle{#1}|}
\def\ket #1{|{#1}\rangle}

\def\beqn{\begin{eqnarray}}
\def\eeqn{\end{eqnarray}}
\def\beq{\begin{equation}}
\def\eeq{\end{equation}}
\def\phan{\vphantom{\frac12}}

\def\non{\nonumber}
\def\al{\alpha}
\def\be{\beta}
\def\ga{\gamma}
\def\Ga{\Gamma}
\def\la{\lambda}
\def\La{\Lambda}

\def\D{{\cal D}}
\def\L{{\cal L}}
\def\de{\delta}
\def\cP{{\cal P}}
\def\({\left(}
\def\){\right)}

\def\om{\omega}
\def\d{\partial}

\author{S. M. Klishevich%
 \thanks{E-mail address: klishevich@mx.ihep.su} \\
         {\it  Institute for High Energy Physics
         } \\
         {\it Protvino, Moscow Region, 142284, Russia
         }
       }
\title{Massive Fields with Arbitrary Integer Spin in Symmetrical Einstein
Space}

\begin{document}
\maketitle
\thispagestyle{empty}
\begin{abstract}
We study the propagation of gauge fields with arbitrary integer spins in
the symmetrical Einstein space of any dimensionality. We reduce the problem
of obtaining a gauge-invariant Lagrangian of integer spin fields in such
background to algebraic problem of search for a set of operators
with certain features by means of the representation of higher-spin fields
in the form of vectors in pseudo-Hilbert space. We consider such
construction at linear order in the Riemann tensor and scalar curvature and
also present an explicit form of interaction Lagrangians and gauge
transformations for massive particles of spins~$1$ and~$2$ in terms of
symmetrical tensor fields.
\end{abstract}
\vskip2mm
\noindent
{\it PACS number(s):} 11.10.Kk, 11.10.Lm, 11.15.-q, 11.15.Ex 
\vskip2mm
\noindent

\newpage
\setcounter{page}{1}

\newpage
\setcounter{page}{1}
\section*{Introduction}

Problems of obtaining a consistent description of the gravitational
interactions of higher-spin fields have the particular significance since it
allows one to connect the higher-spin fields with the observable world.

It is well known that the gravitational interaction of massless fields with
spins $s\ge2$ does not exist in an asymptotically flat space-time%
~\cite{Aragone:79PL}. For the covariant description of physical fields one
must replace the ordinary derivatives in the Lagrangian and gauge
transformations with the covariant ones. Since the covariant derivatives do
not commute, the gauge invariance fails and a residual appears. For the
fields with spins $s\ge2$ the residual is proportional to the Riemann
tensor. In general, one cannot cancel such a residual in an asymptotically
flat space-time in linear approximation by any changes of the Lagrangian and
transformations. Therefore, in such case this approximation does not
exist. Since linear approximation does not depend on the presence
of any other fields in the system, this means that the whole theory of
interaction does not exist either.

This problem can be overcome in several ways. For instance, one can
consider the massless fields in a space of constant curvature. In this case
the Lagrangian for gravity would have the additional term
$\de\L\sim\sqrt{-g}\la$, where $\la$ is the cosmological constant.
A modification of the Lagrangian and the transformations leads to a mixing
of terms with different numbers of the derivatives. This allows one to
compensate the residual with terms proportional to $R_{\mu\nu\al\be}$. The
complete theory will be represented as series in inverse value of the
cosmological constant \cite{Vas-DS-1,Vas-DS-2}. This means the
non-analyticity of the theory in $\la$ at zero, i.e. the impossibility of a
smooth transition to the flat space. Such a theory was considered in
Refs.~\cite{Vas-DS-1,Vas-DS-2,Vasilev:IJMP95}.

Besides, massive higher-spin fields can have the gravitational interaction.
For example, the string theory represents the consistent gravitational
interaction of the massive higher-spin fields. In Ref.~\cite{Porrati} the
interaction of the fields at linear order was derived while investigating
three point functions ot the type II superstring, which contains one
graviton and two massive states.

In the literature the gravitational interaction of arbitrary spin fields
were considered at the lowest order in the Riemann tensor
\cite{P-93,CDP-94,Porrati}. Considering the interactions, the authors
started from the free theory of the massive fields in the conventional
form \cite{Sing-Hagen}. The "minimal" introduction of the interaction leads
to contradictions, therefore, it is necessary to consider non-minimal terms
in the interaction Lagrangian. Since, the conventional
Lagrangian~\cite{Sing-Hagen} for the massive spin-$s$ fields is not gauge
invariant, in this approach there are no restrictions on the form of the
non-minimal interaction. But, in a general case, such a theory is
pathological, therefore, to build a consistent theory of the interaction it
is necessary to introduce an additional restrictions on the non-minimal
terms. So, for instance, when investigating the
gravitational interactions \cite{P-93,CDP-94}, the authors required for the
theory to have the tree-level unitarity up to the Planck scale.
 
In our opinion it is more convenient to use the gauge-invariant approach
when one analyzes an interaction of the massive fields \cite{Zinoviev-83,%
Zinoviev-2,mass_spin}, \cite{Pashnev-1} or \cite{Argyres}. Under such an
approach the interaction is considered as a deformation of initial gauge
algebra and Lagrangian\footnote{Of course, one must consider only a
non-trivial deformation of the free algebra and Lagrangian which cannot
be completely gauged away or removed by a redefinition of the fields.}
\cite{Fronsdal-2,mass_spin}. Although, generally speaking, the gauge
invariance does not ensure the consistency of massive theories, but, 
anyway, it allows one to narrow the searches and conserves the appropriate
number of physical degree of freedom. Besides, such an approach is quite
convenient and practical.

Here we go along the line of Refs.~\cite{oscB,HSferm} where the
electromagnetic interaction of the massive fields with integer and
half-integer spins was investigated. We represent a free state with the
arbitrary integer spin $s$ as state $\ket{\Phi^s}$ in a Pseudo-Hilbert
space\footnote{The representation of the free fields with arbitrary integer
spins in such a form was considered in Refs.
\cite{Labastida,Pashnev-1,Pashnev:MPL97}}
\cite{oscB}. The tensor fields corresponding to the particle with spin $s$
are coefficient functions of the state $\ket{\Phi^s}$. On the considered
Fock space we introduce a set of operators by means of which we define
the gauge transformations and necessary constraints for the state
$\ket{\Phi^s}$. The gauge-invariant Lagrangian has the form of the
expectation value of the Hermitian operator, which consists of the
operators, in the state~$\ket{\Phi^s}$.

In the considered approach the gauge invariance is a consequence of
commutation relations of the introduced operators. For the covariant
description of fields in the Riemann background, one must replace the
ordinary derivatives with the covariant ones. This leads to a change of
algebraic features of the operators and, as a consequence, to the loss of
the gauge invariance for the higher-spin fields. We reduce the problem of
recovering the invariance to algebraic problem of search for such modified
operators that depend on the Riemann tensor and scalar curvature and satisfy
the same commutation relations as initial operators in the flat space. In
this, we should note that in the massless case one cannot realize such a
construction in an asymptotically flat space. In section~3, for the massive
theory we construct the set of the operators having the algebraic features
of the free ones at linear order in the Riemann tensor and scalar curvature.
Besides, in the next section we give an explicit form of interaction
Lagrangian and transformations for the massive vector and spin-2 fields.

\section{Free Field with Spin $s$}
\label{Free}
\subparagraph{Massless fields.}
 Let us consider the Fock space generated by the creation and annihilation
operators $\bar a_\mu$ and $a_\mu$, which are vectors on the $D$-dimensional
Minkowski space ${\cal M}_D$ and which satisfy the following algebra
\begin{eqnarray}
\label{H-alg}
\left[a_\mu,\bar a_\nu\right]&=&g_{\mu\nu}, \quad 
a_\mu^{\dag} = \bar a_\mu,
\end{eqnarray}
where $g_{\mu\nu}$ is the metric tensor with the signature
$\|g_{\mu\nu}\|={\rm diag}(-1,1,1,...,1)$. Since the metric is indefinite,
the Fock space that realizes the representation of Heisenberg algebra
\reff{H-alg} is Pseudo-Hilbert.
 
Let us consider the state in the introduced space:
\begin{equation}
\label{Fstate}
\ket{\Phi^s}=\frac1{\sqrt{s}}\Phi_{\mu_1\dots\mu_s}(x)
\prod_{i=1}^s\bar a_{\mu_i}\ket{0}.
\end{equation}
Coefficient function $\Phi_{\mu_1\dots\mu_s}(x)$ is a symmetrical tensor
field of rank $s$ on space ${\cal M}_D$. For this tensor field to describe
the state with spin\footnote{We consider symmetric tensor fields only.} $s$ 
one has to imposes the condition:
\begin{equation}
\label{2traceless}
\Phi_{\mu\mu\nu\nu\mu_4...\mu_s}=0.
\end{equation}
In terms of such fields the Lagrangian \cite{Fronsdal-1,Curt1} has the form
\begin{eqnarray}
\label{LB-free}
\L_s&\!=\!&\frac12(\partial_\mu\Phi^s)
  \cdot(\partial_\mu\Phi^s)
  -\frac{s}{2}(\partial\cdot\Phi^s)\cdot(\partial\cdot\Phi^s)
  -\frac{s(s-1)}4(\partial_\mu\Phi'^s)
   \cdot(\partial_\mu \Phi'^s)  
\non\\&&{}\!\!\!\!
  -\frac{s(s-1)}{2}(\partial\cdot\partial\cdot\Phi^s)\cdot\Phi'^s
  -\frac18s(s-1)(s-2)(\partial\cdot\Phi'^s)
   \cdot(\partial\cdot\Phi'^s).
\end{eqnarray}
The following notation $\Phi'=\Phi_{\mu\mu...}$ have been used here while
the point means the contraction of all
indexes $\Phi^s\cdot\Phi^s=\Phi_{\mu_1\ldots\mu_s}\Phi^{\mu_1\ldots\mu_s}$.

This Lagrangian is invariant under the transformations
\begin{eqnarray}
\label{Gauge-0}
\de\Phi_{\mu_1\dots\mu_s}&=&\partial_{(\mu_1}\La_{\mu_2\dots\mu_{s-1})}, 
\\\label{tr-less}
\La_{\mu\mu\mu_3...\mu_{s-1}}&=&0.
\end{eqnarray}

Let us introduce the following operators on our pseudo-Hilbert space
\begin{equation}
\label{L-operator0}
L_1=p\cdot a,\quad L_{-1}=L_1^{\dag},\quad
L_2=\frac12a\cdot a,\quad L_{-2}=L_2^{\dag},\quad L_0=p^2.
\end{equation}
Here $p_\mu=i\partial_\mu$ is the momentum operator that acts on the space
of the coefficient functions. 

Operators of type \reff{L-operator0} appear as constraints of a two-particle
system under quantization\footnote{It is also possible to regard operators
\reff{L-operator0} as a truncation of the Virasoro algebra.}
\cite{Barducci}. This operators satisfy the commutation
relations:
\begin{equation}
\label{L-algebra0}
 \begin{array}{rclrcl}
 \left[L_1,L_{-2}\right]&=&L_{-1},
&\quad\left[L_1,L_2\right]&=&0,\\
 \left[L_2,L_{-2}\right]&=& N + \frac{D}{2},
&\quad\left[L_0,L_n\right]&=&0,\\
 \left[L_1,L_{-1}\right]&=&L_0,
&\quad\left[N,L_n\right]&=&{}-nL_n,\quad n=0,\pm 1,\pm 2.
 \end{array}
\end{equation}
Here $N=\bar a\cdot a$ is the level operator that defines the spin of
states. So, for instance, for state~\reff{Fstate}
$$
N\ket{\Phi^s} =s\ket{\Phi^s}.
$$

In terms of operators \reff{L-algebra0} condition \reff{2traceless} can be
written as
\begin{equation}
\label{Trace-L}
(L_2)^2\ket{\Phi^s}=0,
\end{equation}
while gauge transformations \reff{Gauge-0} take the form 
\begin{equation}
\label{Gauge-L}
\de\ket{\Phi^s}=L_{-1}\ket{\La^{s-1}}.
\end{equation}
Here, the gauge state
$$
\ket{\La^{s-1}}=\La_{\mu_1...\mu_{s-1}}\prod_{i=1}^{s-1}\bar a_{\mu_i}
\ket{0}
$$
satisfies the condition
\begin{equation}
\label{traceless}
L_2\ket{\La}=0.
\end{equation}
This condition is equivalent to \reff{tr-less} for the coefficient 
functions.

Lagrangian \reff{LB-free} can be written as the expectation value of a
Hermitian operator in state~\reff{Fstate}
\begin{equation}
\label{Llagr}
\L_s=\bra{\Phi^s}\L(L)\ket{\Phi^s}, \quad \bra{\Phi^s}=\ket{\Phi^s}^{\dag},
\end{equation}
where
\begin{eqnarray}
\label{L-action}
\L(L)&=&L_0-L_{-1}L_1-2L_{-2}L_0L_2-L_{-2}L_{-1}L_1L_2
\non\\&&{}
+\left\{L_{-2}L_1L_1 + h.c.\right\}.
\end{eqnarray}

Lagrangian \reff{Llagr} is invariant under transformations \reff{Gauge-L}
as a consequence of the relation
$$
\L(L)L_{-1}\sim(...)L_2.
$$

\subparagraph{Massive fields}

Let us consider the massive states of arbitrary spin $s$ in the similar
manner. For that we have to extend our Fock space by introducing scalar
creation and annihilation operators $\bar b$ and $b$, which satisfy
the usual commutation relations
\begin{equation}
\label{H-alg+}
\left[b,\bar b\right]=1,\quad b^{\dag}=\bar b.
\end{equation}

Operators \reff{L-operator0} are modified as follows:
\begin{equation}
\label{L-m}
L_1=p\cdot a + mb ,\quad L_2=\frac12\(a\cdot a + b^2\),
\quad L_0=p^2+m^2.
\end{equation}
Here $m$ is an arbitrary parameter having the dimensionality of mass.
In the non-interacting case one can consider such transition as the
dimensional reduction ${\cal M}_{D+1}\to{{\cal M}_D\otimes S^1}$ with the
radius of compactification $R\sim1/m$ (refer also to
\cite{Pashnev-1,Pashnev:MPL97}).

We shall describe the massive field with spin $s$ as the following vector
in the extended Fock space:
\begin{equation}
\label{FockM}
\ket{\Phi^s}=\sum\limits_{n=0}^s
\Phi_{\mu_1\dots\mu_{n}}(x)\bar b^{s-n}\prod_{i=1}^n\bar a_{\mu_i}\ket{0}.
\end{equation}
Like the massless field case, this state satisfies the same condition
\reff{Trace-L}, but in terms of operators \reff{L-m}. The algebra of
operators \reff{L-algebra0} changes insignificantly, the only commutator
modified is
\begin{equation}
\label{comL2m}
 \left[L_2,L_{-2}\right]=N + \frac{D+1}{2}.
\end{equation}
Here, as in the massless case, the operator $N=\bar a\cdot a + \bar bb$
defines the spin of massive states. The Lagrangian describing the massive
field with spin $s$ has the form \reff{L-action} as well, where the
expectation value is taken in state \reff{FockM}. Such Lagrangian is
invariant under transformations \reff{Gauge-L} with the gauge Fock vector
$$
\ket{\La^{s-1}}=\sum\limits_{n=0}^{s-1}\La_{\mu_1...\mu_n}
\bar b^{s-n-1}\prod_{i=1}^n\bar a_{\mu_i}\ket{0},
$$
which satisfies condition \reff{tr-less}.

\section{Propagation of Massive higher-spin Field in Symmertical Einstein
Space}
\label{Inter}
In this section we consider an arbitrary $D$-dimensional symmetrical
Einstein space, i.e. the Riemann space defined by the following equations:
\beqn
\D^{\(\Ga\)}_\mu R_{\nu\al\be\ga} &=& 0,
\\
R_{\mu\nu} -\frac12g_{\mu\nu}R &=& g_{\mu\nu}\la,
\eeqn
where $\D^\Ga_\mu$ is the covariant derivative with the Cristofel
connection
$\Ga^\al{}_{\nu\mu}$. We assume that the Greek indexes are global while the
Latin ones are local. As usual, the derivative $\D^\Ga_\mu$ acts on tensor
fields with global indexes only.

To describe the massive higher-spin fields in the Riemann background, we
must replace the ordinary derivatives with the covariant one, i.e. we make
the substitution:
\beq\label{covariant}
p_\mu\to\cP_\mu=i\(\D^\Ga_\mu+\omega_\mu{}^{ab}\bar a_aa_b\),
\eeq
where $\om_\mu^{ab}$ is the Lorentz connection. We imply that the creation
and annihilation operators primordially carry the local indexes. We also
have to introduce the non-degenerate vielbein $e^a_\mu$ for the transition
from the local indexes to the global ones and vice versa. As usual, we
impose the conventional requirement on the vielbein
$$
\D_\mu^{\(\Ga+\om\)} e^a_\nu =
\d_\mu e_\nu^a - \Ga^\la{}_{\nu\mu}e_\la^a + \om_\mu{}^a{}_be^b_\nu=0.
$$
By means of this relation one can transfer from expressions with one
connection to those with other. Besides, we should note that due to this
relation the operator $\cP_\mu$ commutes with the vector
creation-annihilation operators carrying global indexes 
$\bar a_\nu=e_\nu^b\,\bar a_b$ and $a_\nu=e_\nu^b\,a_b$. This allows us not
to care about the ordering of operators \reff{L-m}.

 One can verify that the covariant momentum operator defined in this way
 properly acts on the states of type \reff{FockM}, indeed
$$
\cP_\mu\ket{\Phi}=i\D_\mu^{\(\om\)}\Phi^{b_1\dots b_n}
\prod_{i=1}^n\bar a_{b_i}\ket{0}
 = i\D_\mu^{\(\Ga\)}\Phi^{\nu_1\dots\nu_n}\prod_{i=1}^n\bar
a_{\nu_i}\ket{0}.
$$

 The commutator of the covariant momenta defines the Riemann tensor:
\begin{equation}
\left[\cP_\mu,\cP_\nu\right]=R_{\mu\nu}{}^{ab}\(\om\)\bar a_aa_b.
\end{equation}
where $R_{\mu\nu}{}^{ab}\(\om\)=\d_\mu\om_\nu{}^{ab}
+\om_\mu{}^a{}_c\,\om_\nu{}^{cb} - \(\mu\leftrightarrow\nu\)$.

In the definition of operators \reff{L-m}, we replace the ordinary momenta
with the covariant ones as well. As a result, the operators cease to obey
algebra~\reff{L-algebra0}. Therefore, Lagrangian \reff{L-action} loses the
invariance under gauge transformations~\reff{Gauge-L}.

To recover the gauge invariance, we do not need to restore total
algebra~\reff{L-algebra0}, it is enough to ensures the existence of the
following commutation relations:
\beqn
\label{needR1}
[L_1,L_{-1}] &=& L_0,
\\
\label{needR2}
 [L_2,L_{-1}] &=& L_1.
\eeqn

To restore these relations, let us represent operators \reff{L-m} as 
normal ordered functions of the creation and annihilation operators as well
as of $R_{\mu\nu\al\be}$ and $R$, i.e.
$$
L_i=L_i\(\bar a_\mu,\bar b,a_\mu,b,R_{\mu\nu\al\be},R\).
$$
The particular form of the operators $L_i$ is defined from the condition
recovering of commutation relations \reff{needR1} and \reff{needR2} by
these operators. We should note that it is enough to define the form of 
the operators $L_1$ and $L_2$, since the operators $L_0$ and $N$ can be
expressed in terms of these operators.

Since we have turned to the extended universal enveloping algebra of the
Heisenberg algebra, the arbitrariness in the definition of the operators $a$
and $b$ appears. Besides, we should admit the presence of arbitrary operator
functions depending on $a$, $b$, $R_{\mu\nu}{}^{ab}$, and $R$ in the
right-hand side of \reff{H-alg} and \reff{H-alg+}. In this, such a
modification of the operators must not lead to breaking the Jacobi
identity and under the transition to flat space they must restore the
initial algebra. However, one can make sure that using the arbitrariness in
the definition of the creation and annihilation operators, we can restore
algebra \reff{H-alg}, \reff{H-alg+} at linear order in the Riemann tensor
and scalar curvature. 

We shall search for the operators $L_1$ and $L_2$ as series in the Riemann
tensor and scalar curvature.

Let us consider linear approximation.

Operator $L_1$ should be no higher than linear in the operator $\cP_\mu$,
since the presence of a greater number of these operator changes type of the
gauge transformations and the number of physical degrees of freedom.
Therefore, in this approximation we can search for them in the form
\beqn\label{L1anzR}
L_1^{(1)}&=&
	 R\biggl(h_0(\bar b,b)\,b 
 + h_1(\bar b,b)\,b\( \bar a\cdot a\)
 +  \bar b h_2(\bar b,b)\,a^2 + h_3(\bar b,b)\,b^3\bar a^2
\non\\&&{}
 + h_4(\bar b,b)\(\cP\cdot a\) + h_5(\bar b,b)\,b^2\(\bar a\cdot\cP\)
\biggr)
	 +  R^{\mu\nu ab}\biggl(
	   h_6(\bar b,b)\, b\bar a_\mu\bar a_a a_\nu a_b
\non\\&&{}
	 + h_7(\bar b,b)\,\bar a_\mu a_\nu \cP_a a_b
	 + h_8(\bar b,b)\, b_2\bar a_\mu\cP_\nu\bar a_a a_b\biggr).
\eeqn
At the same time the operator $L_2$ cannot depend on the momentum operators
at all, since condition \reff{Trace-L} defines the purely algebraic
constraint on the coefficient functions. Therefore, at this order we choose
the operator $L_2$ in the following form:
\beqn\label{L2anzR}
L_2^{(1)}&=&
	  R\( h_9(\bar b,b)\, b^2
	+ h_{10}(\bar b,b)\,a^2
	+ h_{11}(\bar b,b) b^2\(\bar a\cdot a\)
	+ h_{12}(\bar b,b)\, b^4\bar a^2\)
\non\\&&{}
	+ h_{13}(\bar b,b)\,b^2 \bar a^\mu\bar a^a a^\nu a^b R_{\mu\nu ab}.
\eeqn
Here $h_i(\bar b,b)$ are normal ordered operator functions
$$
h_i(\bar b,b)=\sum\limits_{n=0}^\infty H_n^i \bar b^n b^n,
$$
where $H^i_n$ are arbitrary real coefficients.

Let us define a particular form of the functions $h_i$ from the condition
of
recovering commutation relations~\reff{needR1} and~\reff{needR2} by the
operators~$L_1$ and~$L_2$. 

We have to note that these operators can obey relations~\reff{needR1}
and~\reff{needR2} up to the terms proportional to
$L_2^{(0)}=\frac12\(a^2+b^2\)$ at linear order, since this does not
break the gauge invariance due to constraint~\reff{traceless}.
 
Having calculated \reff{needR1} and passing to normal symbols of the
creation and annihilation operators, we obtain a system of differential
equations in the normal symbols of operator functions $h_i$. For the
normal symbols of operator functions we shall use the same notations. This
does not lead to the mess since we consider the operator functions as the
ones of two variables, while their normal symbols as functions of
one variable. Thereby, we have equations from \reff{needR1}
\begin{eqnarray}
\label{Dif_eq1}
&& h_{7}''(x) + 2 h_{7}'(x) + 4 h_{13}(x) - 2 h_{8}(x) = 0,
\non\phan\\&&
x\(h_{6}''(x)+2 h_{13}'(x)+2 h_{6}'(x)\)+2\(h_{6}'(x)+2 h_{13}(x)\) = 0,
\non\phan\\&&
x^2\(\frac{1}{2}h_{8}''(x)+h_{8}'(x)\)+2 x \(h_{8}'(x)+h_{8}(x)\)
 + h_{8}(x) - 2 h_{7}(x) = 0,
\non\phan\\&&
\(h_{2}''(x) + 2 h_{12}'(x) x + 2 h_{2}'(x) + 8 h_{12}(x) - 2 h_{3}(x)\)=0,
\non\phan\\&&
h_{4}''(x) + 2 h_{4}'(x) + 2 h_{11}(x) - 2 h_{5}(x) = 0,
\non\phan\\&&
x\(\frac{1}{2} h_{1}''(x) + h_{11}'(x) + h_{1}'(x)\) + h_{1}'(x) + 2
h_{11}(x) + 2 h_{2}(x) = 0,
\non\phan\\&&
x^2\(\frac{1}{2} h_{5}''(x) + h_{5}'(x)\) + 2 x \(h_{5}'(x) + h_{5}(x)\)
 + 2 h_{10}(x) + \frac1{D}h_{7}(x) + h_{5}(x) = 0,
\non\phan\\&&
x^3\(\frac{1}{2}h_{3}''(x) + h_{3}'(x)\) + 3 x^2 \(h_{3}'(x) + h_{3}(x)\)
 + x\( - \frac{1}{2} h_{0}''(x) + h_{10}'(x)
\non\phan\right.\\&&\left.{}
 - h_{9}'(x) - h_{0}'(x)
 + \frac1{D}h_{6}(x) + 3 h_{3}(x) - h_{2}(x) + h_{1}(x)\)
\non\phan\\&&{}
 - h_{0}'(x) - 2 h_{9}(x) - Dh_{2}(x)=0.
\end{eqnarray}
Here the prime denotes the derivative with respect to $x$, while
$x=\bar\be\be$, where $\bar\be$ and $\be$ are the normal symbols of the
operators $\bar b$ and $b$, correspondingly.

Similarly, from \reff{needR2} we derive the other system of equations:
\begin{eqnarray}
\label{Dif_eq2}
&& h_8'(x) =0,
\non\phan\\&&
h_8''(x) x + h_7''(x)  + 3 h_8'(x)  + 2 h_6'(x) =0,
\non\phan\\&&
h_8'(x) x + 3 h_7'(x)  + 2 h_8(x) + 2 h_6(x) = 0,
\non\phan\\&&
x^2 h_8''(x) + x\(h_7''(x) + 4 h_8'(x) + 6 h_6''(x)\)
 + h_7'(x) + 2h_8 + 6h_6 - 4 = 0,
\non\phan\\&&
h_6''(x) x + 2 h_6'(x) =0,
\non\phan\\&&
h_8'(x) x +\frac12 h_7'(x)  + 2 h_8(x) + h_6(x) = 0,
\non\phan\\&&
x\(h_5''(x) + 2 h_3'(x)\) + h_4''(x)  + 3 h_5'(x) + 4 h_2'(x) + h_1'(x)
 + 6 h_3(x) = 0,
\non\phan\\&&
h_1''(x) x + 2 h_1'(x) = 0,
\non\phan\\&&
h_5'(x) x + 3 h_4'(x)  + 2 h_5(x) + 2 h_2(x) + h_1(x) = 0,
\non\phan\\&&
h_5''(x) x^2 + \(h_4''(x) + 4 h_5'(x) + 2 h_2'(x) + 3 h_1'(x)\)x
 + h_4'(x) + 2 h_5(x)
\non\phan\\&&{}
 + 2 h_2(x) + 3 h_1(x) = 0,
\non\phan\\&&
h_3''(x) x^2 + \(h_2''(x) + 6 h_3'(x)\)x + 2 h_2'(x) + 6 h_3(x) = 0,
\non\phan\\\phan&&
h_3''(x) x^3 + \(h_2''(x) + 4 h_3'(x)\)x^2 - 2 h_0''(x) x
 - 4 h_0'(x)=0.
\end{eqnarray}
Having solved the whole system\footnote{We search for regular at $x\to0$
solutions only.} of equations \reff{Dif_eq1} and \reff{Dif_eq2}, we
obtain the particular form of the operators $L_1$ and~$L_2$:
\beqn\label{Resh_i}
L_1^{(1)}&=& 
\frac{1}{6}R^{\mu\nu\al\be}\bar\al_\al\al_\mu
\left\{\cP_\nu\al_\be\(1+2\bar\be\be\) - 5\bar\al_\nu\al_\be \be
 + 2\bar\al_\nu\cP_\be \be^2\right\}
\non\\&&{}
 + R\left\{
 c_2\(\cP\cdot\al\) + c_1\be
 - \frac{1}{2}\al^2\bar\be^2 h_5'(x)\be
 - \al^2 \bar\be h_5(x)
\right.\non\\&&\left.{}
 + \(\bar\al\cdot\cP\)h_5(x)\be^2
 + \frac{1}{2} \bar\al^2 h_5'(x)\be^3
\right\},
\non\\
L_2^{(1)}&=&
 R\left\{\al^2\(- \frac{1}{4}\bar\be^2h_5''(x)\be^2
 - \frac{1}{2}\bar\be^2h_5'(x)\be^2
 - \bar\be h_5'(x)\be - \bar\be h_5(x)\be
\right.\right.\non\\&&\left.{}
 - \frac{1}{2} h_5(x)
 + \frac1{6D} \bar\be\be + \frac{1}{12D}\)
 + \frac{1}{4}\bar\al^2\( h_5''(x)+ 2 h_5'(x)\)\be^4
\non\\&&\left.{}
 + \(\bar\al\cdot\al\)h_5\be^2
 + \frac{1}{2}  h_5(x)\be^2 D + \frac{1}{3D}\bar\be\be^3\right\},
\eeqn
where $c_1$ and $c_2$ are arbitrary real parameters and $h_5(x)$ is an
arbitrary function regular at $x\to 0$, while $\bar\al_\mu$ and $\al_\mu$
are normal symbols of the operators $\bar a_\mu$ and $a_\mu$. One can
verify that the function $h_5(x)$ corresponds to the rest of the
arbitrariness in the redefinition of the creation and annihilation operators
when initial Heisenberg algebra \reff{H-alg}, \reff{H-alg+} is fixed.
Therefore, we can set $h_5(x)=0$.

The transition to the operator functions is realized in the conventional
manner:
$$\left.
:\!O(\bar a,\bar b,a,b)\!:=
\exp{\(\bar a\cdot\frac{\partial}{\partial\bar\al}\)}
\exp{\(\bar b\frac{\partial}{\partial\bar\be}\)}
\exp{\(a\cdot\frac{\partial}{\partial\al}\)}
\exp{\(b\frac{\partial}{\partial \be}\)}
 O(\bar\al,\bar\be,\al,\be)
\right|_{{\al^\#\to0}\atop{\be^\#\to0}}.
$$

 Thus, we have obtained the general form of the operators $L_n$, which
satisfy commutation relations \reff{needR1} and \reff{needR2} in linear
approximation. This means that Lagrangian \reff{L-action} is invariant under
gauge transformations \reff{Gauge-L} at this order. The form of the operator
$L_2$ has changed in this approximation, hence, the conditions
$$
L_2L_2\ket{\Phi^s}=0, \qquad L_2\ket{\La^{s-1}}=0
$$
undergo the nontrivial modifications in terms of the coefficient functions.

\section{Examples}
In this section we apply the proposed algebraic scheme to the
description of propagation of the massive states with spin $1$ and $2$ in
the Symmetrical Einstein space.

\subparagraph{Vector massive field.} This case is quite interesting
since it is practically the only massive bosonic field among the other
higher-spin states which was observed in the experiment. Let us consider the
state in the Hilbert space that corresponds to the massive state with 
spin~$1$.
$$
\ket{1}=\(\(v\cdot\bar a\)+\varphi\bar b\)\ket{0}.
$$

It is not difficult to compute the expectation value of operator
\reff{L-action} in this state. Having calculate this, one derive, the
following Lagrangian\footnote{We suppress the usual factor $\sqrt{-g}$.} in
linear approximation 
\begin{eqnarray}\label{LinearS1}
\L_{s=1} &=&
\(1+2c_2R\)\(\bar v^\al \cP^2 v_\al
-\bar v^\be \cP_\be \cP_\al v^\al
+\bar\varphi \cP^2 \varphi\)
+\(1+2c_1R\)\bar v^\al v_\al
\non\\&&{}
-\(1+(c_1+c_2)R\)\(\bar\varphi \cP_\al v^\al+ h.c. \)
+\frac13\bar v_\de R^{\al\de\ga\be}\cP_\al \cP_\ga v_\be.
\end{eqnarray}
For the Lagrangian to describe the massive vector field properly, we
have to impose the constraints
\beq\label{Gconstraints}
1+2c_1R\ge0,\qquad 1+2c_2R>0.
\eeq
The former constraint ensures the given state not to be the tachyon, while
the latter one provides kinetic terms with the right sign.

The gauge vector for the massive spin-1 state is
$$
\ket{\La,1}=\eta b\ket{0}
$$
and the gauge transformations for the massive field are
\begin{eqnarray*}
\de v_\al &=& \(1 + c_2 R\) \cP_\al\eta,
\\
\de\varphi &=& \(1 + c_1 R\)\eta.
\end{eqnarray*}

For the vector massive field it is not difficult to generalize the linear
approximation to the general case\footnote{Of course, this is only one
possibility among others.} of arbitrary symmetrical Einstein space.
For that we make the following substitution:
$$
c_1R\to f_1(R), \qquad c_2R\to f_2(R).
$$
But the gauge invariance requires the functions to be equal to each
other. Thereby, the whole Lagrangian describing the propagation of the
vector field in the considered background is
\begin{eqnarray}
\label{ToyS1}
\L_{s=1}\! &\!=\!&
\(1+f(R)\)\(\bar v^\al \cP^2 v_\al
-\bar v^\be \cP_\be \cP_\al v^\al
+\bar\varphi \cP^2 \varphi
+\bar v^\al v_\al
-\(\bar\varphi \cP_\al v^\al+h.c. \)\)
\non\\&&{}
+\frac13\bar v_\de R^{\al\de\ga\be}\cP_\al \cP_\ga v_\be.
\end{eqnarray}
There is no reason to be surprised, since, due to the gauge invariance, we
cannot obtain a different result by virtue of the fact that $R$ is
constant.

Now we can consider two variants. The first is when
$1+f(R)>0$\footnote{The massive vector field becomes a ghost when
$1+f(R)<0$.}. Then, the Lagrangian is invariant under the usual gauge
transformations for the massive vector fields
\begin{eqnarray*}
\de v_\mu &=&\cP_\mu\eta,
\\
\de\varphi &=& \eta.
\end{eqnarray*}
In principle, in this case we can include the multiplier $1+f(R)$ into
the normalization of fields. After that we obtain the usual Lagrangian
"minimally" coupled to the Riemann background with the single "non-minimal"
term. 

A different quite unusual situation is realized when $1+f(R)=0$. Then, the
whole Lagrangian consists of the single term
$$
\L={}\frac13R^{\mu\nu\al\be}\bar V_{\mu\nu}V_{\al\be},
$$
where $V_{\mu\nu} = \d_\mu v_\nu - \d_\nu v_\mu$. Obviously, the
Lagrangian is invariant under the transformation $\de v_\mu=\cP_\mu\eta$.
One can notice that the transition to an arbitrary Riemann space does not
break this invariance.

Let us discuss the causality for the massive vector field in the given
background. Having fixed the gauge invariance by
$$
\varphi=0,
$$
from \reff{LinearS1} we derive the following equations
$$
\frac{\de\L}{\de\bar v^\mu}
=\(1+2c_2R\)\(\cP^2v_\mu - \cP_\mu\(\cP\cdot v\)\)+\(1+2c_1R\)v_\mu
-\frac13R_\mu{}^{\nu\al\be}\cP_\nu\cP_\al v_\be.
$$
Having taken the divergence of these equations one obtains the constraint
$$
\cP^\mu\frac{\de\L}{\de\bar v^\mu}=
\(1+\(2\(c_1-c_2\)-\frac1D\)R\)\(\cP\cdot v\) + {\cal O}\(R^2\)=0.
$$
From here we see that if we impose the requirement
\beq\label{ConstrS1a}
m^2+\(2\(c_1-c_2\)-\frac1D\)R\neq0
\eeq
 we get the necessary constraint on the vector field at this order. Here we
have restored the dimensional parameter $m$. Calculating the characteristic
determinant\footnote{The determinant is entirely determined by the
coefficients at the highest derivatives in equations of motion after
gauge fixing and resolving all the constaraints \cite{Velo1:69}.}, we
obtain as a result
$$
D(n)=\({3D}m^2+R\)\(n^2\)^D + {\cal O}(R^2),
$$
where $n_\mu$ is the normal vector to the characteristic surface.
The equations of motion will be causal (hyperbolic) if the solutions $n^0$
to $D(n)=0$ are real for any $\vec n$. Thereby, in our case, from the
condition $D(n)=0$ we have the usual light cone as the solution for $n_\mu$
if we impose the following condition:
\beq\label{ConstrS1b}
{3D}m^2+R\neq0
\eeq
 Of course, our consideration essentially depends on the higher orders in
the Riemann tensor and scalar curvature.

Thus, we can see that in such theory there are restrictions on $m$ and
$R$ similar to Ref.~\cite{Higuchi:NP87,Bengtsson-95}.

\subparagraph{Massive field with spin $2$.}
Now we obtain the Lagrangian describing the propagation of massive spin-2
field in the symmetrical Einstein space. The following state in the Fock
space corresponds to such field
$$
\ket{2}=\{\(\bar a\cdot h\cdot \bar a\)
+\(v\cdot\bar a\)\bar b + \varphi b^2
\}\ket{0}.
$$
It is easy to see that this state trivially satisfies
condition\footnote{One can verify that condition \reff{Trace-L} imposes a
not-trivial restriction only on the states with spin $4$ and higher.}
\reff{Trace-L}.

Having calculated the expectation value of operator \reff{L-action} in
this state we derive the following Lagrangian
\begin{eqnarray}\label{RLagrS2}
\!\!\L_{s=2}&\!\!=\!\!&
\(1+2c_2R\)\Biggl\{\bar h^{\al\be}\cP^2h_{\al\be}-2\bar h\cP^2h
-2\bar h^{\be\ga}\cP_\be\cP_\al h^\al_\ga
+\biggl\{2\bar h\cP_\al \cP_\be h^{\al\be}
\non\\\phan&&{} 
 +\bar \varphi\cP_\al\cP_\be h^{\al\be}
-\bar h\cP^2\varphi
+h.c.\biggr\}
+\frac12\bar v^\al \cP^2v_\al
 -\frac12\bar v^\be \cP_\al \cP_\be v^\al
\Biggr\}
\non\\\phan&&{} 
 -\(1+\(c_1+c_2\)R\)\left\{\bar v_\be \cP_\al h^{\al\be} 
 - \bar h\cP_\al v^\al
 +h.c.\right\}
\non\\\phan&&{} 
+\(1+2c_1R\)\(\bar h^{\al \be }h_{\al\be }-\bar hh\)
-\frac38R^{\al\ga\be\de}\bar h_{\ga \mu }\cP_\al\cP_\de h_\be^\mu
\non\\\phan&&{}
-\frac12R^{\al\ga\be\de}\bar v_\ga\cP_\al\cP_\de v_\be
-\frac16\biggl\{ 
 2R^{\al\ga\be\de}\bar h_{\be \ga}\cP_\al\cP_\mu h_\de^\mu
-R^{\al\ga\be\de}\bar h_{\be \ga}\cP_\al \cP_\de h
\non\\\phan&&{}
-5R^{\al\ga\be\de}\bar h_{\be \ga }\cP_\al\cP_\de\varphi
+5R^{\al\ga\be\de}\bar h_{\be \ga }\cP_\al v_\de
+h.c.\biggr\} 
+\frac43R^{\al\ga\be\de}\bar h_{\al\be}h_{\ga\de}
\non\\\phan&&{}
+\frac1{3D}R\Biggl\{ {}
-\bar h\cP^2h 
+\left\{\frac12\bar h\cP_\al\cP_\be h^{\al\be}
- \frac12\bar h\cP^2\varphi
 + h.c.\right\}
+3 \bar\varphi\cP^2\varphi
\non\\\phan&&{}
+\left\{\frac12\bar v^\al\cP_\al h
 -3 \bar \varphi \cP_\al v^\al
+h.c.\right\}
-4 \bar hh 
+3\bar v^\al v_\al\Biggr\}
\end{eqnarray}
where $h=g^{\mu\nu}h_{\mu\nu}$. It is not difficult to notice that we have
to impose the same restrictions~\reff{Gconstraints} for the proper
description of the massive state.

The gauge vector for the massive spin-2 state is 
$$
\ket{\La,2}=\{\(\xi\cdot\bar a\) + \eta b\}\ket{0}.
$$
Condition \reff{traceless} is the non-trivial constraint for the gauge
vectors of massive states with spin $3$ and higher only.

From \reff{Gauge-L} we obtain the following gauge transformations
\begin{eqnarray*}
\de h_{\al \be }&=&\(1 + c_2R\)\cP_{(\al }\xi _{\be )}
-\frac16 R_{(\al}{}^\ga{}_{\be)}{}^\de\cP_\ga\xi_\de, 
\\\phan
 \de v_\al &=&\(1 + c_2R\)\cP_\al u+\(1+c_1R\)\xi _\al, 
\\\phan
\de \varphi &=&\(1+c_1R\)\eta.
\end{eqnarray*}

Let us fix the gauge invariance by means of the gauge condition
$$
v_\mu =0,\qquad \varphi=0.
$$
Now we have the following equations of motion
\beqn\label{S2eq}
\frac{\de \L}{\de h^{\mu\nu}}&=&
\(1+2c_2R\)\biggl(\cP^2 h_{\mu\nu}
 - 2\cP_{(\mu}\(\cP\cdot h\)_{\nu)} + \cP_{(\nu}\cP_{\mu)}h
\non\\&&{}
- g_{\mu\nu}\(\cP^2h-\(\cP\cdot\cP\cdot h\)\)\biggr)
 - \(1+2c_1R\)\(h_{\mu\nu} - g_{\mu\nu}h\)
\non\\&&{}
 + \frac{R}{6D}\(\cP_{(\nu}\cP_{\mu)}h
 - g_{\mu\nu}\left\{
 2 \cP^2h
 - \(\cP\cdot\cP\cdot h\)
 - 8 h \right\}\)
\non\\&&{}
 - \frac{1}{6} g_{\mu\nu} R^{\al\de\be\ga} \cP_\de\cP_\ga h_{\al\be}
 + \frac{1}{3} R_{\al(\mu|\be}{}^\ga \cP_{|\nu)}\cP_\ga h^{\al\be}
 - \frac{1}{6} R^\be{}_{(\mu}{}^\ga{}_{\nu)} \cP_\ga\cP_\be h
\non\\&&{}
 + \frac{1}{3} R_{\be(\mu}{}^\ga{}_{\nu)} \cP_\ga \(\cP\cdot h\)^\be
 - \frac{2}{3} R^{\al\ga\be}{}_{(\mu|} \cP_\ga\cP_\be h_{|\nu)\al}
 - \frac{4}{3} R_{\al\mu\be\nu} h^{\al\be}.
\eeqn
From these equations we can obtain the constraints:
\beqn\label{S2contr}
\non
&&\(\cP\cdot h\)_\mu= \cP_\mu h \(1 + \frac{R}{6D}\)
 + \frac56 R_\mu{}^{\nu\al\be}\cP_\al h_{\nu\be},
\\&&{}
\(3 + 2 \(3 c_1 D + \frac2{D}\) R\)h=0.
\eeqn
Now one can see that when
$$
3 + 2 \(3 c_1 D + \frac2{D}\) R\neq0
$$ 
we have the appropriate number of the constraints and, correspondingly, the
appropriate number of physical degree of freedom at this order.

Now we consider the question of causality of the massive spin-2 state in 
linear approximation. Using relations \reff{S2contr} and the equations of
motion \reff{S2eq}, one can obtain the characteristic determinant
$$
D(n)=\(n^2\)^{\frac{D^2(D+1)^2}{4}}\(1+\frac{D+5}{3D}R\) + {\cal O}\(R^2\).
$$
Thereby, when $\(1+\frac{D+5}{3D}R\)\neq0$ from the condition $D(n)=0$, we
have the usual light cone in this approximation, i.e. the causal
propagation of the massive spin-2 state in the considered Riemann
background.

For the massive states with higher spins, one can derive a similar result.
Quite obviously that the causal propagation of these states in considered
background imposes some restrictions on the mass of states and
the scalar curvature only.

\section{Conclusion}
We have applied the algebraic scheme proposed in
Ref.~\cite{oscB,HSferm} to the description of propagation of the gauge
massive fields in the arbitrary symmetrical Einstein space of arbitrary
dimensionality in the lowest approximation in the Riemann tensor and scalar
curvature. This approach is quite convenient since it allows one to reduce
the cumbersome problem of searching for the gauge invariant action of
higher-spin fields to the pure algebraic problem of search for the
appropriate modification of some operators. In principal, this approach can
be applied to the description of fermionic massive fields in such background
as well.

\section*{Acknowledgment}

The author is grateful to Yu. M. Zinoviev for the numerous
useful discussions and the help in the work.


\end{document}